# Fault Detection and Classification using Wavelet and ANN in DFIG and TCSC Connected Transmission Line

Satya Vikram Pratap Singh[1*], Tanu Prasad[1†], Siddharth Kamila[1†] and Prashant Agnihotri[1†]

[1]Department of Electrical Engineering and Computer Science, Indian Institute of Technology (I.I.T.) Bhilai, Near G.E.C. Campus, Sejbahar, Raipur, 492015, Chhattisgarh, India.

*Corresponding author(s). E-mail(s): satyavp@iitbhilai.ac.in;
Contributing authors: tanuprasad12345@gmail.com; siddharthk@iitbhilai.ac.in; pagnihotri@iitbhilai.ac.in;
†These authors contributed equally to this work.

**Abstract**

This paper presents fault detection and classification using Wavelet and ANN-based methods in a DFIG-based series compensated system. The state-of-the-art methods include Wavelet transform, Fourier transform, and Wavelet-neuro fuzzy methods-based system for fault detection and classification. However, the accuracy of these state-of-the-art methods diminishes during variable conditions such as changes in wind speed, high impedance faults, and the changes in the series compensation level. Specifically, in Wavelet transform-based methods, the threshold values need to be adapted based on the variable field conditions. To solve this problem, this paper has proposed a Wavelet-ANN-based fault detection method where Wavelet is used as an identifier and ANN is used as a classifier for detecting various fault cases. This methodology is also effective under SSR condition. The proposed methodology is evaluated on various fault and non-fault cases generated on an IEEE first benchmark model under varying compensation levels from **20%** to **55%**, impedance faults, and wind velocity from 6m/sec to 10m/sec using MATLAB/Simulink, OPAL-RT(OP4510) manufactured real-time digital simulator environment, Arduino board I/O ports communicating with external PC in which ANN model dumped, using Arduino support package of MATLAB. The preliminary results are compared with the state-of-the-art fault detection method, where the proposed method shows robust performance under varying field conditions.

# 1 Introduction

In Electrical Power System Networks, fixed series compensation is used to increase the power transfer capabilities. But the series compensated transmission lines with a DFIG-based wind farm produces the condition of Sub-Synchronous Resonance (SSR) at high compensation levels and low wind speed, which can be mitigated using FACTS devices such as Thyristor Controlled Series Compensator (TCSC). SSR is the condition in which DFIG exchanges energy with a TCSC-connected electrical network at sub-synchronous or super-synchronous frequency. Due to SSR, the phase currents undergo amplitude and phase distortion, producing oscillation and leading to system instability. Under these unstable conditions, fault identification and classification become difficult and result in false detection [1 – 4].

Faulty conditions in transmission lines need to perform three major tasks i.e., fault detection, fault classification, and finding fault location. The fault must be detected accurately and rapidly to protect the system from harmful effects. Transmission line faults are classified as series (open conductor) fault and shunt (short circuit) fault. The short circuit fault occurs more frequently and can be identified by monitoring the current of each phase. Whenever a short circuit fault occurs, the current value of the respective phase increases. Different algorithms have been proposed for fault detection [5] and classification, such as prominent, hybrid, and modern techniques [6,7]. Prominent techniques are based on Signal Processing Techniques (SPT) like Wavelet transforms, Fourier transforms, S-transform, artificial neural network, and the Fuzzy-logic approach [8–12]. To overcome the drawback of detection time and complexity in prominent techniques, the hybrid technique is used, which is the integration of any two prominent techniques. Some of the hybrid techniques are Neuro-fuzzy technique, Wavelet-neuro-fuzzy technique, Wavelet-artificial neural network technique, Wavelet-fuzzy logic technique [13 – 15]. Now the era of intelligent protection systems is gaining popularity which detects, classify, and locate faults in less time and is accurate, but complex. Modern techniques include support vector machine, genetic algorithm, DWT- ELM approach, FPGA-based implementation, GSM technique, PMU-based protection scheme, decision tree-based method, and multi-information measurement [16 – 23]. Apart from the many features of these techniques, the priority is fast fault detection and classification. A comprehensive comparison of fault detection techniques in Table I, shows the high complexity level for existing techniques. Techniques suffer from a lack of satisfactory results under consideration of varying field conditions such as fault resistance, compensation level, and wind speed. Assigning varying Threshold values for fault detection under varying field condition is also difficult in existing techniques.

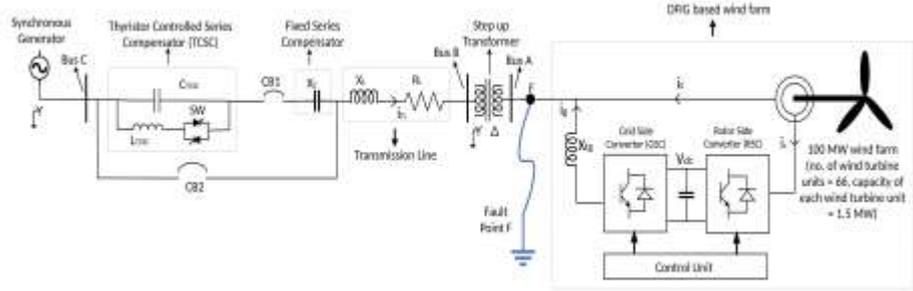

Fig.1 Schematic Representation of a 100 MVA, 60 Hz, IEEE first benchmark power system

This paper proposes a method for fault detection and classification in transmission lines using Wavelet and ANN. The Wavelet Transforms give the maximum value of detailed coefficients of all three phases and the ground. During the fault condition, the coefficient of the respective faulty phases is higher than the non-faulty phases. ANN is trained based on all the healthy and fault conditions considering different compensation levels, wind speeds, and fault resistances value. Whenever a fault occurs, the Wavelet transform provides the maximum coefficient, which acts as input for the ANN model, and at the end, it classifies the faulty phases. Compared to the previous pre-existing technique, wavelet transformation does not provide accurate results in different compensation levels, wind speed, and fault resistance [2]. The ANN avoids the problem of setting a threshold value under varying conditions of fault resistance ranging from 0.01Ω to 0.1Ω, compensation level range from 20% to 60%, and the wind speed range from 6 m/sec to 10 m/sec. This range of fault resistance is taken to study high fault current. During Online mode, the output of wavelet data is input to that of the ANN model, which successfully detects the fault. In the proposed technique, the complexity level and Fault detection time is less as compare to existing techniques. Also problem of threshold value for fault detection under varying field condition does not exists. This paper has five sections. The System description is outlined in section II. Then, the detailed description of Wavelet and ANN-based fault detection and classification are in section III. Then, the simulation and hardware results are highlighted in section IV. Finally, the paper ends with the conclusion in section V.

**Table I** Comparison of Fault Detection Techniques

| Approach | Technique | Complexity |
|---|---|---|
| **Prominent Techniques** | | |
| Wavelet Approach | Wavelet Transform, DWT | Medium |
| Artificial Neural Network Approach | ANN, Distributed and hierarchical NN (DHNN), back-propagation (BP) | Complex |
| Fuzzy Logic Approach | Fuzzy set approach | Simple |
| **Hybrid Techniques** | | |
| Neuro-Fuzzy Technique | Neural network, fuzzy logic, Fuzzy neural network (FNN), ANFIS | Complex |
| Wavelet and ANN Technique | DWT, CWT, ANN | Medium |
| Wavelet and Fuzzy logic Technique | Fuzzy set approach | Simple |
| Wavelet and neuro-fuzzy Technique | Neural network, fuzzy logic, wavelet transform | Complex |

| Modern Techniques | | |
|---|---|---|
| Support Vector Machines | SVM Classifier, wavelet | Complex |
| Genetic Algorithm | GA, NN | Complex |
| DWT-ELM Approach | DWT, ELM | Medium |
| Wavelet and neuro-fuzzy Technique | Neural network, fuzzy logic, wavelet transform | Medium |
| Decision Tree based Method | Discrete Fourier Transform | Complex |
| Pattern recognition approach | Multi resolution Analysis | Complex |

## 2 System Description

For the study of fault detection and classification, IEEE first benchmark model has been used, shown as a single-line diagram in Fig.1 and system parameters are shown in Table II. The wavelet and ANN block have been added to the Simulink model to detect and classify faults in online mode. The model consists of a 100 MVA, 60 Hz Synchronous Generator connected to Bus C with a voltage level of 161kV and a DFIG-based wind farm connected to Bus A with a voltage level of 575V. The step-up transformer used is of delta-star configuration, which steps up the voltage from 575V to 161kV. The fixed series compensator which is represented here by a series capacitor is used in the transmission line to increase the steady-state power limit of the system. But it causes instability in the system by creating the condition of Sub-Synchronous Resonance (SSR). Thyristor-Controlled Series Compensator (TCSC) improves stability by mitigating SSR. The fault created at a point F shown in Fig.1 in the duration of 4.5-4.55 sec for software results and, intentionally created through external switch for hardware results, to detect and classify the fault.

## 3 Methods

In the proposed methodology, Wavelet is used to identify the fault, and ANN is used to classify the fault. Wavelet is implemented using the code written in the function block of Simulink and the ANN model are trained using the ANN toolbox of MATLAB. The detection and classification have been done in online mode. This section discusses the flow of the process, discrete wavelet transforms, description of the ANN model, difficulty in fault detection due to SSR, and variation of threshold under varying field conditions.

### 3.1 Process Flow Chart

Firstly, the current from each phase and ground is given to the Wavelet code implemented in the function block for the feature extraction. It decomposes the signal into approximate (*cA*) and detailed coefficient (*cD*) using *db*4

**Table II** Parameters for the IEEE first benchmark power system used in Simulation

| Sl.No. | Parameters | Values |
|---|---|---|
| 1. | Rated Power | 100 MVA |
| 2. | Frequency | 60 Hz |
| 3. | Wind Turbine Speed | 6 to 10 m/sec |
| 4. | Voltage Generated by DFIG | 575 V |
| 5. | Voltage between DFIG and Bus A | 575 V |
| 6. | Voltage between Bus A and Bus B | 161 KV |

| 7. | Voltage between Bus B and Bus C | 161 KV |
|---|---|---|
| 8. | Voltage between Bus C and Synchronous Generator | 161 KV |
| 9. | Step up transformer | 100 MVA, 575 V/161 KV, 60 Hz |
| 10. | Transmission line between Bus B and Grid | Positive sequence parameters: - $R_1$ = 5.1842$ohm$, $X_1$ = 129.605$ohm$ |
|  |  | Zero sequence parameters: - $R_0$ = 15.5526$ohm$, $X_0$ = 388.815$ohm$ |
| 11. | Series Capacitor Compensator | $X_c$ = 129.605$ohm$, K = 20% |
| 12. | Thyristor Controlled Series Compensator (TCSC) | $L_{TCSC}$ = 0.0123$H$, $C_{TCSC}$ = 90.9629µ$F$, alpha = 68 degrees |
| 13. | Synchronous Generator | 100 MVA, 60 Hz |

Wavelet with layer one decomposition, then the maximum of detailed coefficient (*max*(*cD*)) calculated using the same wavelet code. The *db*4 Wavelet with layer one decomposition is taken based on hit and trail method considering various wavelets and different layers of decomposition. Threshold value is taken as maximum value of (*max*(*cD*)). This threshold value is used to train the ANN model, which classifies the fault condition with logic 1 and no-fault condition with logic 0.

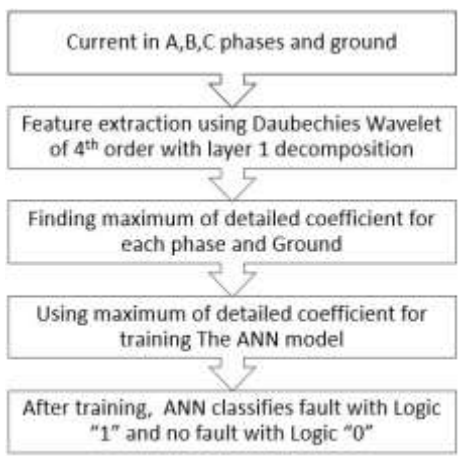

Fig.2 Flow chart for fault detection and classification

## 3.2 Discrete wavelet Transform

In the proposed methodology, the Daubechies Wavelet of order four (*db*4) with single layer decomposition has been used to detect the fault in the given transmission line network. Here, with layer one decomposition, $cD_1$ and $cA_1$ specify the detailed and approximate coefficients.

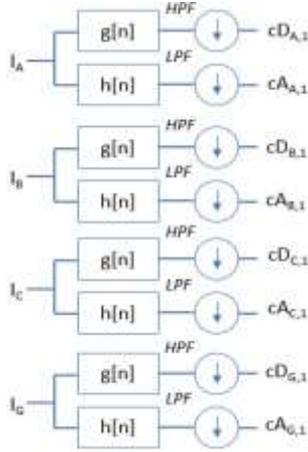

Fig.3 Wavelet Decomposition tree of phase currents and ground

$$\psi_{a,b}(t) = \frac{1}{\sqrt{a}} \psi\left(\frac{t-b}{a}\right); a > 0, b \in \Re \quad (1)$$

The translation and dilation parameters are shown by *a* and *b*, respectively. To obtain the DWT, parameters a and b need to be discretized. Discretizing $a = 2^j$ and $b = k2^j$ will yield orthonormal basis functions for certain choices of $\psi$ (db Wavelets) [24].

$$\psi_{(j,k)}(t) = 2^{-j/2} \psi(2^{-j}t - k) \quad (2)$$

Multi Resolution Analysis (MRA) can be used to obtain the DWT of a discrete signal by applying low-pass and high-pass filters, iteratively, and subsequently down sampling them by two. Fig.3 illustrates this process, where g[n] and h[n] are the high-pass and low-pass filters, respectively, g[n] denotes the mother wavelet and h[n] denotes scaling function [24]. x[n] denotes the phase or ground currents. At each level, this procedure computes,

$$y_{high}[k] = \sum_n x[n] \cdot g[2k - n] \quad (3)$$

$$y_{low}[k] = \sum_n x[n] \cdot h[2k - n] \quad (4)$$

where $h[N - 1 - n] = (-1)^n g[n]$ with N being the total number of samples in x[n] and $y_{high}[k]$ and $y_{low}[k]$ are the outputs of high pass and low pass filters, respectively, at each level. The number of levels this process is repeated depends on the choice of the user. At the last level, the $y_{low}[k]$ obtained is called as approximate and $y_{high}[k]$ computed at each level is called as the detailed coefficient at that level [24]. The values of *m, n, p* and *q* is given as :

$$m = max(cD_{A,1}) \quad (5)$$

$$n = max(cD_{B,1}) \qquad (6)$$

$$p = max(cD_{C,1}) \qquad (7)$$

$$q = max(cD_{G,1}) \qquad (8)$$

Table III Dataset of Fault Coefficients during AG Fault Condition at $V_w$ = 6m/sec

| Sl.No. | Compensation Level (in %) | Fault Resistance (in ohms) | m | n | p | q |
|---|---|---|---|---|---|---|
| 1. | 20 | 0.01 | 144.6 | 1.421e-4 | 1.431e-4 | 144.6 |
|  |  | 0.1 | 15.84 | 1.555e-5 | 1.563e-5 | 15.84 |
| 2. | 30 | 0.01 | 143.9 | 1.414e-4 | 1.416e-4 | 144 |
|  |  | 0.1 | 15.79 | 1.551e-5 | 1.557e-5 | 15.79 |
| 3. | 40 | 0.01 | 141.3 | 1.382e-4 | 1.379e-4 | 141.3 |
|  |  | 0.1 | 15.52 | 1.525e-5 | 1.52e-5 | 15.53 |
| 4. | 50 | 0.01 | 142.5 | 1.404e-4 | 1.412e-4 | 142.5 |
|  |  | 0.1 | 15.59 | 1.534e-5 | 1.539e-5 | 15.59 |
| 5. | 60 | 0.01 | 187.8 | 2.763e-4 | 1.027e-4 | 187.8 |
|  |  | 0.1 | 20.61 | 1.126e-4 | 6.467e-5 | 20.62 |

Table IV Dataset of Fault Coefficients during AB Fault Condition at $V_w$ = 6m/sec

| Sl.No. | Compensation Level (in %) | Fault Resistance (in ohms) | m | n | p | q |
|---|---|---|---|---|---|---|
| 1. | 20 | 0.01 | 160.7 | 160.7 | 1.185e-3 | 1.179e-7 |
|  |  | 0.1 | 16.1 | 16.1 | 7.54e-5 | 7.476e-08 |
| 2. | 30 | 0.01 | 161.6 | 161.6 | 1.122e-4 | 1.086e-07 |
|  |  | 0.1 | 16.13 | 16.13 | 6.834e-5 | 6.782e-8 |
| 3. | 40 | 0.01 | 161.6 | 161.6 | 9.724e-4 | 9.51e-9 |
|  |  | 0.1 | 16.16 | 16.16 | 7.499e-5 | 7.391e-9 |
| 4. | 50 | 0.01 | 159.7 | 159.7 | 8.903e-5 | 8.87e-9 |
|  |  | 0.1 | 15.96 | 15.96 | 5.192e-5 | 5.076e-9 |
| 5. | 60 | 0.01 | 118.9 | 118.9 | 1.38e-4 | 1.34e-8 |
|  |  | 0.1 | 13.48 | 13.48 | 6.491e-4 | 7.293e-8 |

Table V Fault Classification in Binary Form as 0 and 1 according to different fault types

| Sl.No. | Types of Faults | Phase A | Phase B | Phase C | Ground G |
|---|---|---|---|---|---|
| 1. | AG | 1 | 0 | 0 | 1 |
| 2. | BG | 0 | 1 | 0 | 1 |
| 3. | CG | 0 | 0 | 1 | 1 |
| 4. | AB | 1 | 1 | 0 | 0 |
| 5. | BC | 0 | 1 | 1 | 0 |
| 6. | CA | 1 | 0 | 1 | 0 |
| 7. | ABG | 1 | 1 | 0 | 1 |
| 8. | BCG | 0 | 1 | 1 | 1 |
| 9. | CAG | 1 | 0 | 1 | 1 |
| 10. | ABC | 1 | 1 | 1 | 0 |
| 11. | ABCG | 1 | 1 | 1 | 1 |
| 12. | No Fault | 0 | 0 | 0 | 0 |

## 3.3 ANN model description

The neural network toolbox of MATLAB has been used for training the multilayer feed-forward artificial neural network using the Levenberg-Marquardt backpropagation algorithm. The input in this model is the maximum of wavelet coefficients *max(m)*, max(n), *max(p)*, and *max(q)*, which is the data of *max(cD)* calculated by varying compensation level (K) of TCSC from 20% to 60%,with a gap of 5%,wind velocity of DFIG- based wind farm from 6*m/sec* to 10*m/sec* with a gap of 1*m/sec* and fault resistance from 0.01Ω to 0.1Ω with a gap of 0.01Ω. In Table III, the maximum of fault coefficients for AG fault have been shown for $V_w$ = 6*m/sec*. The output is predicted as logic 0 for no fault condition and logic 1 for the faulted state, for different types of faults as shown in Table IV. The predicted result is shown in four running windows, three for phases and one for ground as shown in Table IV. ANN model comprised an input layer with four neurons, two hidden layers with fifty neurons, and an output layer with four neurons. Activation functions considered are tansig in the input layer, tansig in both hidden layers, and logsig for the output layer. The model successfully predicts the faulty phase with mean square error (MSE) of $10^{-12}$.

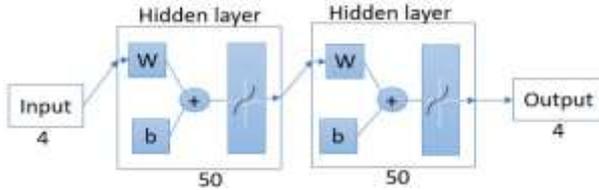

Fig.4 ANN Architecture

## 3.4 Fixed series compensation and effect of SSR

$$SSSL = \frac{V_1 V_2}{X_L(1-K)}; K = \frac{X_c}{X_L}$$

(9)

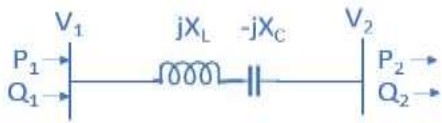

Fig.5 Fixed series compensation

The fixed series capacitor increases the power system network's steady state stability limit (SSSL) with a fixed compensation level(K) [25]. In Fixed series compensated networks, if it is connected with the DFIG-based wind farm, the condition of sub-synchronous resonance (SSR) occurs. SSR is a condition in which a mechanical system exchanges energy with the electrical system at one or more frequencies. During SSR, the positive sequence sub-synchronous current flowing in the generator stator causes oscillations in generator rotor at frequencies 60 + $\omega_n$ and 60 − $\omega_n$. $\omega_n$ is the SSR frequency, and Ω is the rotor oscillation frequency. At Ω = 60−$\omega_n$, oscillations do not damp out in shaft torsional response. During SSR, $\omega_n$ = 60 − Ω [25].

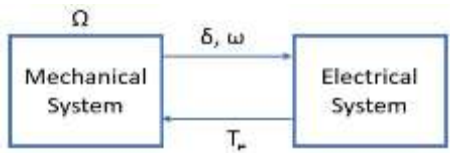

Fig.6 Coupled mechanical and electrical system

$$\omega_n = \frac{1}{\sqrt{LC}}$$
(10)

Considering the isolated mechanical system and isolated electrical system given in Fig.6 in which $\delta$ depict the angle between the generator's internal and terminal voltage and Te depict the electro-mechanical torque. The Mechanical system is affected by Te generated as an effect of the electrical system. $\delta$ and $\omega$ are given as input to the electrical system, which carries the influence of Ω [26]. This effect can be seen in Te which further enhances the mechanical oscillations. Due to the presence of multiple frequency components, the system becomes oscillatory, and the amplitude of phase current gets distorted. Fig.7 shows the phase currents without DFIG connected system and Fig.8 show the phase A current with DFIG connected system under AG and no-fault condition. This distorted and oscillatory system, in the presence of fault, makes fault detection difficult.

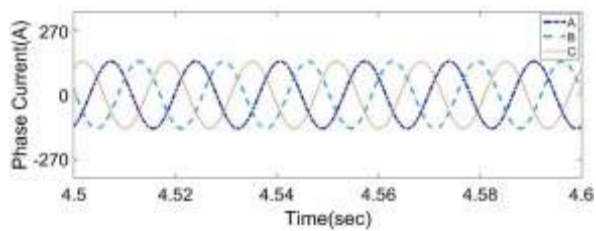

Fig.7 Phase current without DFIG system

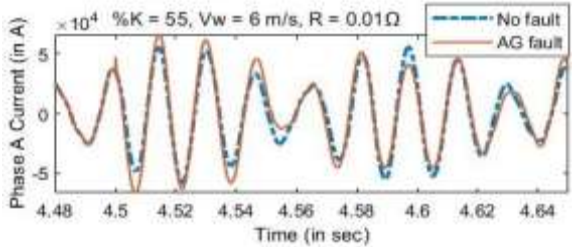

Fig.8 Phase A current with DFIG system at k = 55% R = 0.01 ohm, and wind velocity = 6m/sec

## 3.5 Variation of threshold limits under varying conditions of K, R and $V_w$

A study of variation of threshold values under varying field conditions of K, R, and $V_w$ for AG Fault is shown from fig 9 to fig 12. Fig.9 shows the Maximum of Fault coefficient (m) for AG fault at fault resistances from 0.01 Ω, to 0.1 Ω, under varying compensation levels. The problem of varying value of fault coefficients due to various field conditions such as compensation level (K), fault resistance (R) and wind velocity ($V_w$), makes it difficult to determine a fixed threshold limit of fault coefficients m, n, p, and q. The ANN uses all variable threshold limits as input data and successfully classifies the fault.

Fig 10 shows fault coefficient (m) for AG fault condition under varying conditions of wind velocities ($V_w$) from 6m/s to 10m/s. Fig 11 shows fault coefficient (m) for AG fault condition under varying condition of compensation level from 20% to 60%. From this it is observed that the fault coefficient(m) is very close to (max(m)). Fig 12 shows Fault coefficient (m) for AG fault condition under varying condition of fault resistances. As the coefficient keeps on changing for every instant. So, based on the above observations with consideration of all parameters, maximum of the wavelet coefficient is chosen for training, using ANN.

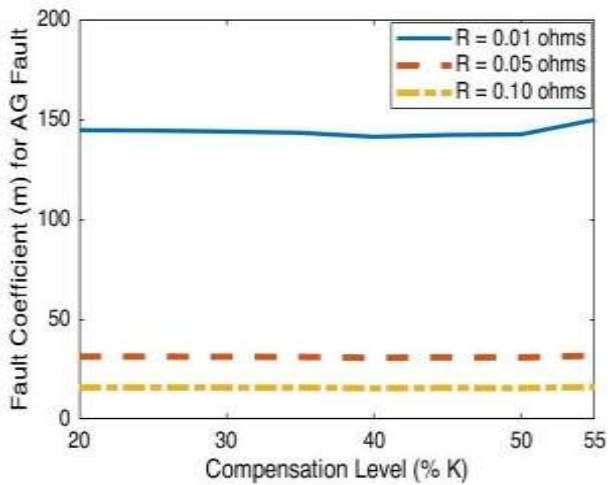

Fig.9 Fault coefficient (m) for AG fault condition with varying fault resistance values

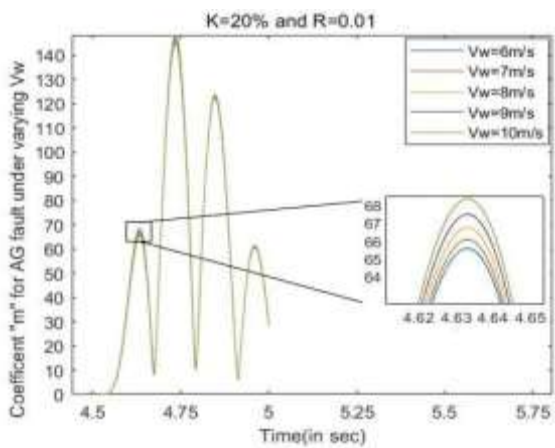

Fig.10 Fault coefficient (m) for AG fault condition with varying condition of wind velocities

## 4 Results and Discussion

### 4.1 Simulation Results

This section shows a comparative study of systems having ANN as a classifier and ANN, not as a classifier for AG, AB, and ABC fault. Detection and classification for all symmetrical and unsymmetrical faults have been carried out in MATLAB/ Simulink. Successful classification of faults through 0/1 logic has been shown for AG, AB, and ABC faults.

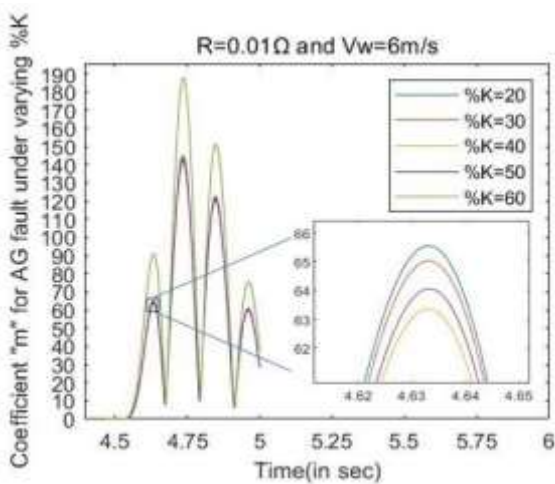

Fig.11 Fault coefficient (m) for AG fault condition with varying condition of compensation level

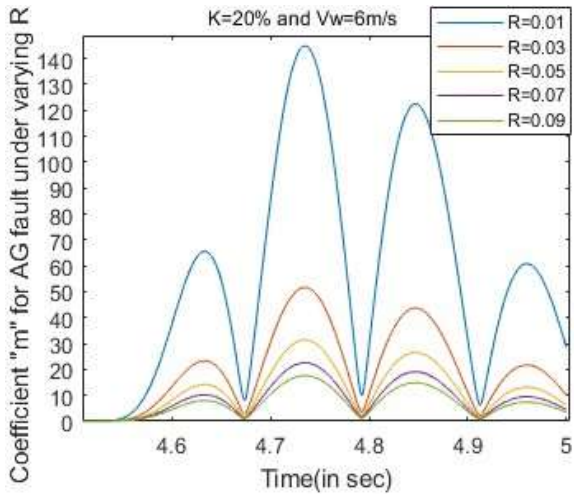

Fig.12 Fault coefficient (m) for AG fault condition with varying condition fault resistance

## 4.2 Software Results of Detection and Classification at different Fault condition

The Detection and Classification at different Fault conditions are done in MATLAB environment. The Fault has been created at 4.5 sec to 4.55 sec for software results. Fig.13 shows Wavelet with ANN and Wavelet without ANN-based AG fault detection.Fig.14 shows Wavelet with ANN and Wavelet without ANN-based AB fault detection.Fig.15 shows Wavelet with ANN and Wavelet without ANN-based ABC fault detection. Results show that Wavelet with ANN successfully predicts all fault conditions with logic 0 under no-fault conditions and logic 1 under faulty conditions.

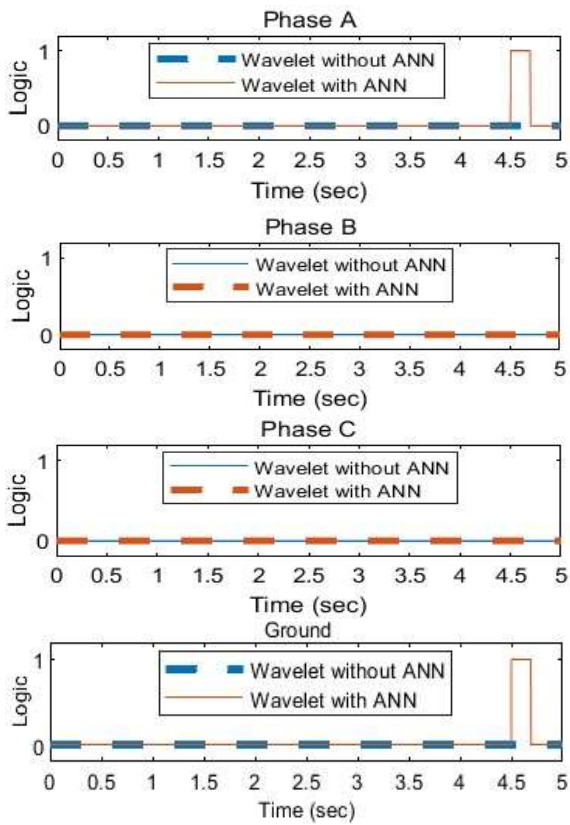

Fig.13 Wavelet with ANN and Wavelet without ANN based AG fault detection

### 4.3 Hardware Setup and Results

Fig 16 shows the proposed Setup for Wavelet-ANN-based fault detection and classification. The main model is loaded in the OPAL-RT OP4510 Real-time digital simulator connected with the host PC and provides fault coefficients at a sampling rate of 10$\mu sec$. The coefficients are monitored in DSO 1 as shown in Fig.19 and send to the Arduino board as an analog signal. The Arduino board communicates with the external PC using MATLAB Support Package for Arduino. The external PC is used to dump the ANN model and to simulate it for an infinite duration. The analog signal is fed to the Arduino board given as input to this ANN block at a sampling rate of 10$msec$. After successful prediction with a mean square error (MSE) of $10^{-12}$, the Predicted value as 0 and 1 logic is monitored in DSO 2 as digital output as shown in Fig.18.

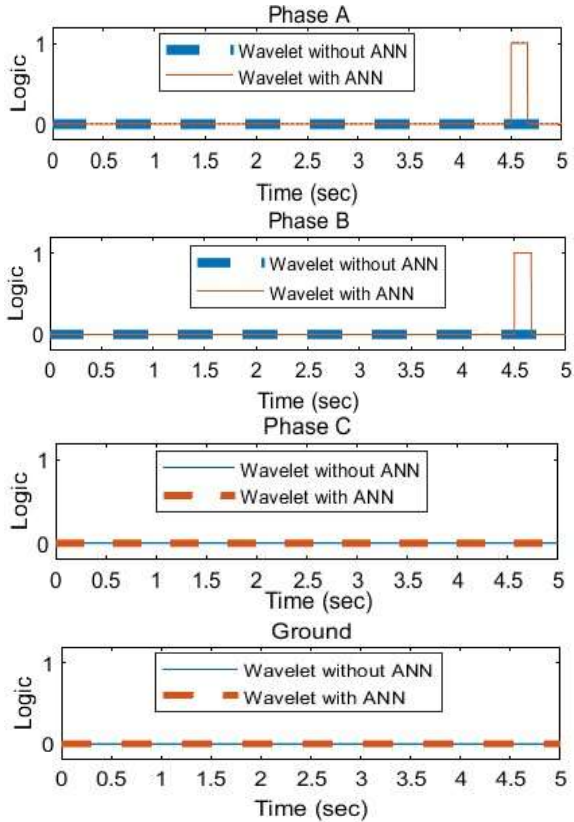

Fig.14 Wavelet with ANN and Wavelet without ANN based AB fault detection

Fig.17 shows Experimental Setup for Power System fault detection using wavelet-ANN in Hardware Synchronised Mode. The controller is load in a real-time digital simulator which is OPAL-RT OP4510 and the real-time results are visualized on the monitoring screen. The Real-time simulation consists of a master and console system. The master system consists of DFIG connected TCSC system where Wavelet Transform is done and fault coefficients are provided to the console system consisting of Scope results of *m*, *n*, *p,* and *q,* also monitored in DSO 1. Along with it, the triggering is provided to create the fault in the system from the console section. The fault coefficient is provided to the ANN model which is dumped in an external PC through the Arduino support package of MATLAB. Arduino is used only in input/output mode, where analog input ports are used for capturing the signal and digital output ports are used to show the predicted output. The fault detection and classification are done for the respective faulty phases in online mode and are monitored at DSO 2. The real-time simulation of

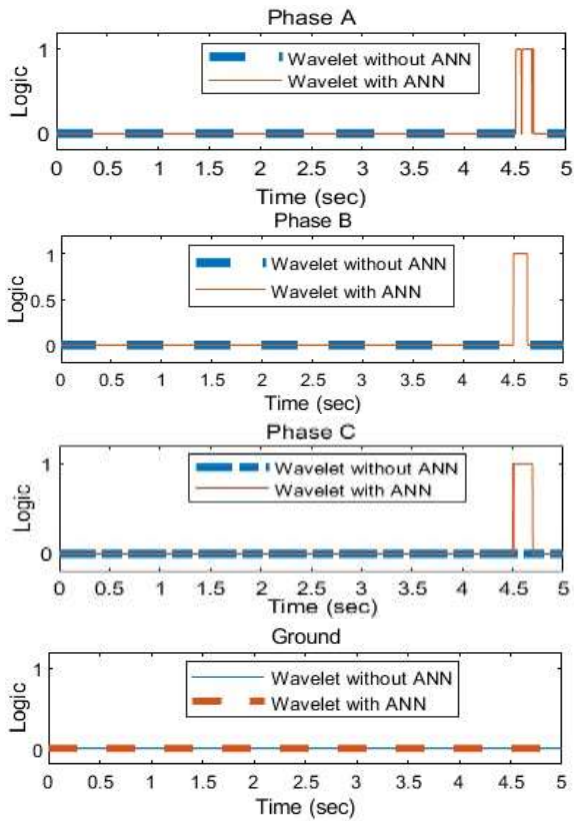

Fig.15 Wavelet with ANN and Wavelet without ANN based ABC fault detection

fault detection and classification are done in hardware synchronized mode. The results of experimental setup shown from Fig.18 to Fig.23. Fig.18 shows the DSO 2 result of Fault Detection and classification for AG fault under varying conditions: - $V_w$ =6m/sec, $K$ =55%, $R$ =0.1Ω in which fault occurs at A phase and Ground with logic 0 under no-fault conditions and logic 1 under faulty conditions is observed. Fig.19 shows the DSO 1 result of Wavelet

Coefficients for AG fault under varying conditions: - $V_w$ =6m/sec, $K$ =55%,

$R$ =0.1 Ω in which fault occurs at A phase and Ground and the magnitude of faulty phase is high. Fig.20 shows the DSO 2 result of Fault detection and classification for AB fault under varying conditions: - $V_w$ =6m/sec, $K$ =20% $R$ =0.01Ω in which fault occurs at A phase and B phase with logic 0 under no-fault conditions and logic 1 under faulty conditions is observed. Fig.21 shows the DSO 2 result of Fault detection and classification for AB fault under varying conditions: - $V_w$ =10m/sec, $K$ =55%, $R$ =0.1Ω. Fig.22 shows the DSO 2 result of Fault detection and classification for ABC fault under varying conditions: - $V_w$ =10m/sec, $K$ =55%, $R$ =0.01Ω in which fault occurs at A phase, B phase and C phase with logic 0 under no-fault conditions and logic 1 under faulty conditions is observed. Fig.23 shows the DSO 2 result of Fault detection and classification for ABC fault under varying conditions: $V_w$ =10m/sec, $K$ =20%, $R$ =0.1Ω in which fault occurs at A phase, B phase and C phase with logic 0 under no-fault conditions and logic 1 under faulty conditions is observed.

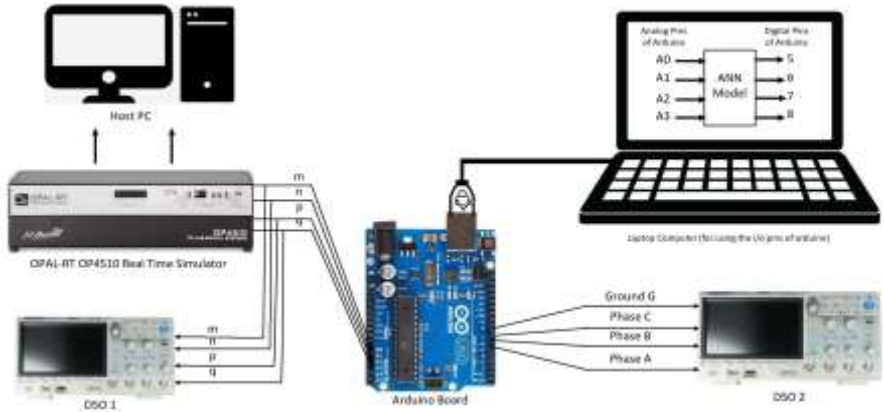

Fig.16 Block Diagram for the wavelet-ANN based power system fault detection

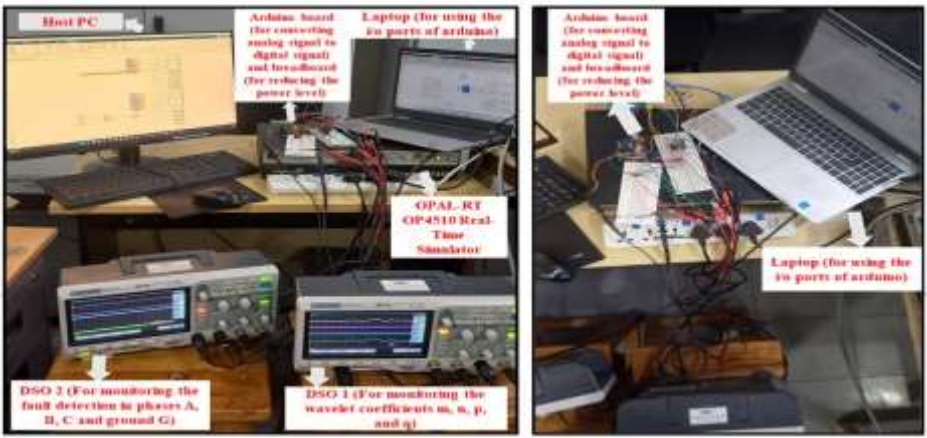

Fig.17 Experimental Setup for Power System fault detection using wavelet-ANN in Hardware Synchronised Mode

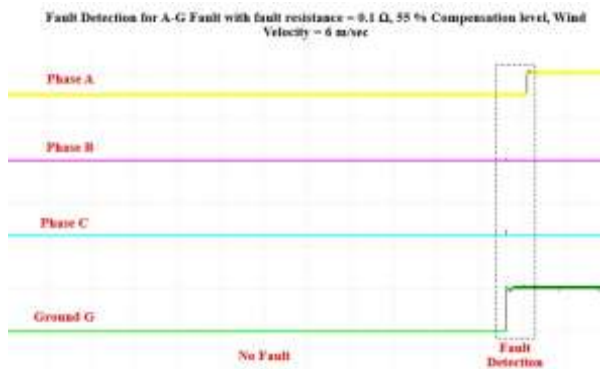

Fig.18 Fault Detection for AG fault with wind speed = 6 m/sec, compensation level = 55 percent, fault resistance = 0.1 ohms

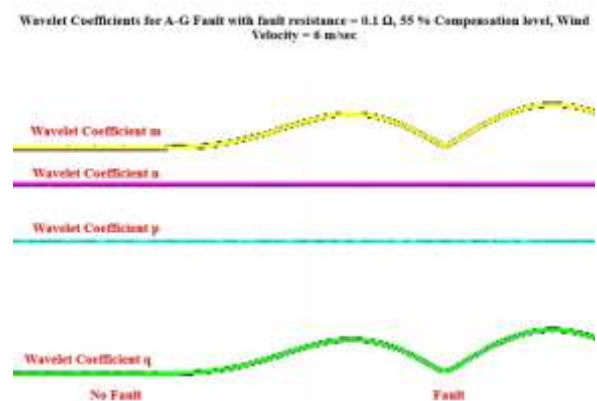

Fig.19 Wavelet Coefficients for AG fault with wind speed = 6 m/sec, compensation level = 55 percent, fault resistance = 0.1 ohms

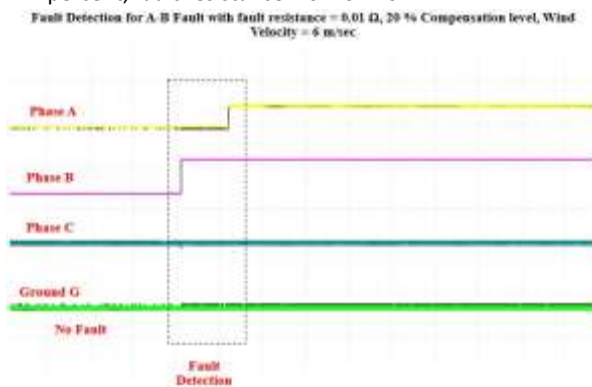

Fig.20 Fault detection for AB fault with wind speed = 6 m/sec, compensation level = 20 percent, fault resistance = 0.01 ohms

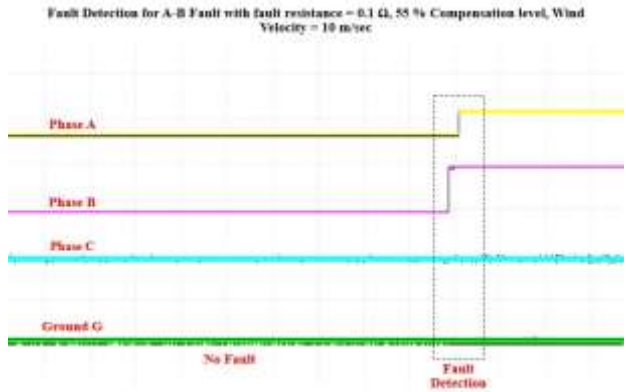

Fig.21 Fault detection for AB fault with wind speed = 10 m/sec, compensation level = 55 percent, fault resistance = 0.1 ohms

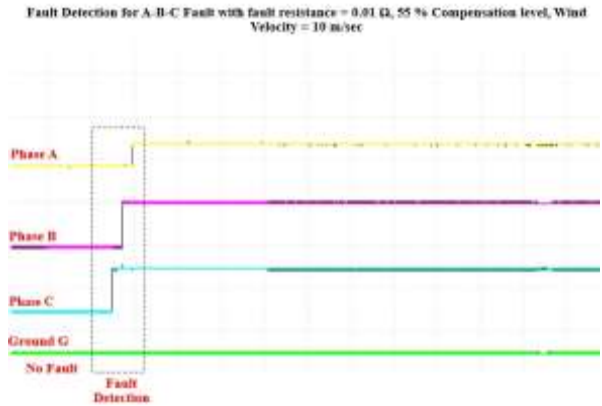

Fig.22 Fault detection for ABC fault with wind speed = 10 m/sec, compensation level = 55 percent, fault resistance = 0.01 ohms

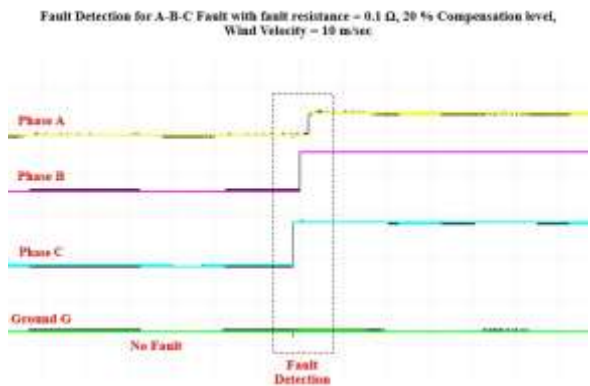

Fig.23 Fault detection for ABC fault with wind speed = 10 m/sec, compensation level = 20 percent, fault resistance = 0.1 ohms

## 5 Conclusions

DFIG connected Transmission lines using Wavelet Transform with ANN and without ANN is investigated in this paper. The software results are achieved in the online mode of MATLAB/Simulink and hardware results are achieved using MATLAB/Simulink, OPAL-RT OP4510 manufactured real-time digital simulator environment, Arduino board I/O ports communicating with external PC in which ANN model dumped, using Arduino support package of MATLAB. Results show that the Wavelet with ANN successfully detects and classify for all short circuit faults in online mode at different wind speeds ($V_w$), compensation levels (K), and fault resistance (R). The following conclusion is made based on results as follows:

1. Accurate detection is achieved with a reduction of mean square error (MSE) to $10^{-12}$ in the case of highly distorted phase and ground currents during the addition of the DFIG system.
2. The Fault detection and classification technique has been done during the online mode.
3. The problem of setting variable thresholds under varying conditions of fault resistance from 0.01Ω to 0.1Ω, compensation level from 20% to 60% and the wind speed range from 6 m/sec to 10 m/sec is eliminated using the ANN model.

## 6 List of abbreviations

ANN: Artificial Neural Network
DFIG: Doubly Fed Induction Generator
MSE: Mean Square Error
PC: Personal Computer
I/O: Input/Output
DSO: Digital storage oscilloscope

TCSC: Thyristor Controlled Series Capacitor

SSR: Sub-Synchronous Resonance
SSSL: Steady State Stability Limit
MRA: Multi Resolution Analysis
DWT: Discrete Wavelet Transform
cD: Detail Coefficient
cA: Approximate Coefficient

## 7 Declarations

### 7.1 Availability of data and materials

Data will be made available on request.

### 7.2 Competing interests

The authors declare no conflict of interest.

### 7.3 Funding

Not applicable

### 7.4 Authors' contributions

Each author has made equal contributions in finding ideas, collecting and analysing data, using software and hardware tools, and writing this paper.

### 7.5 Acknowledgements

Not applicable

## References

[1] L. Wang, X. Xie, Q. Jiang, H. Liu, Y. Li and H. Liu, "Investigation of SSR in Practical DFIG-Based Wind Farms Connected to a Series-Compensated Power System," in IEEE Transactions on Power Systems, vol. 30, no. 5, pp. 2772-2779, Sept. 2015, doi: 10.1109/TPWRS.2014.2365197.

[2] M. Abdeen et al., "Investigation on TCSC Parameters and Control Structure for SSR Damping in DFIG-Based Wind Farm," 2021 12th International Renewable Energy Congress (IREC), 2021, pp. 1-5, doi: 10.1109/IREC52758.2021.9624934.

[3] H. Li et al., "An Improved Fast Detection Method on Subsynchronous Resonance in a Wind Power System With a Series Compensated Transmission Line," in IEEE Access, vol. 7, pp. 61512-61522, 2019, doi: 10.1109/ACCESS.2019.2915545.

[4] H. A. Mohammadpour and E. Santi, "Sub-synchronous resonance analysis in DFIG-based wind farms: Definitions and problem identification — Part I,"

2014 IEEE Energy Conversion Congress and Exposition (ECCE), 2014, pp. 812-819, doi:10.1109/ECCE.2014.6953480.

[5] M. Singh, B. K. Panigrahi and R. P. Maheshwari, "Transmission line fault detection and classification," 2011 International Conference on Emerging Trends in Electrical and Computer Technology, 2011, pp. 15-22, doi: 10.1109/ICETECT.2011.5760084.

[6] S. P. Valsan and K. S. Swarup, "High-Speed Fault Classification in Power Lines: Theory and FPGA-Based Implementation," in IEEE Transactions on Industrial Electronics, vol. 56, no. 5, pp. 1793-1800, May 2009, doi: 10.1109/TIE.2008.2011055.

[7] Avagaddi, Prasad and Edward, Belwin and Ravi, Kuppan. (2018). A Review on Fault Classification Methodologies in Power Transmission Systems: Part I. 5. 10.1016/j.jesit.2017.01.004

[8] A.Swethaa and P. K. Murthy, "Analysis of power system faults in EHVAC line for varying fault time instances using wavelet transforms",Journal of Electrical Systems and Information Technology 2017,Vol-4,pp-107-112

[9] F. B. Costa, B. A. Souza and N. S. D. Brito, "Real-time classification of transmission line faults based on Maximal Overlap Discrete Wavelet Transform," PES TD 2012, 2012, pp. 1-8, doi: 10.1109/TDC.2012.6281684.

[10] S. Kirubadevi and S. Sutha, "Wavelet based transmission line fault identification and classification," 2017 International Conference on Computation of Power, Energy Information and Commuincation (ICCPEIC), 2017, pp. 737-741, doi: 10.1109/ICCPEIC.2017.8290461.

[11] N. Saravanan and A. Rathinam, "A Comparitive Study on ANN Based Fault Location and Classification Technique for Double Circuit Transmission Line," 2012 Fourth International Conference on Computational Intelligence and Communication Networks, 2012, pp. 824-830, doi: 10.1109/CICN.2012.15.

[12] M. R. Bishal, S. Ahmed, N. M. Molla, K. M. Mamun, A. Rahman and M. A. A. Hysam, "ANN Based Fault Detection Classification in Power System Transmission line," 2021 International Conference on Science Contemporary Technologies (ICSCT), 2021, pp. 1-4, doi: 10.1109/ICSCT53883.2021.9642410.

[13] B. Rathore, O. P. Mahela, B. Khan, H. H. Alhelou and P. Siano, "Wavelet-Alienation-Neural-Based Protection Scheme for STATCOM Compensated Transmission Line," in IEEE Transactions on Industrial Informatics, vol. 17, no. 4, pp. 2557-2565, April 2021, doi: 10.1109/TII.2020.3001063

[14] Imtiyaz Alam, M.A.Ansari and Nidhi Singh Pal , "A Comparative Study between Wavelet Primarily Based ANN and ANFIS Algorithm Technique to Locate Fault

in a Transmission Line", Proc. of the 1st IEEE International Conference on Power Electronics, Intelligent Control and Energy Systems, New Delhi, pp. 1-6, July 4-6, 2016

[15] M. Jayabharata Reddy, D.K. Mohanta,A wavelet-fuzzy combined approach for classification and location of transmission line faults,International Journal of Electrical Power Energy Systems,Volume 29, Issue 9,2007,doi: 10.1016/j.ijepes.2007.05.001.

[16] Avagaddi, Prasad and Edward, Belwin and Ravi, Kuppan. (2018). A Review on Fault Classification Methodologies in Power Transmission Systems: Part II. 5. 10.1016/j.jesit.2016.10.003.

[17] M. Egan, J. Thapa and M. Benidris, "Machine Learning Using High Precision Data for Fault Location," 2022 17th International Conference on Probabilistic Methods Applied to Power Systems (PMAPS), 2022, pp. 1-5, doi: 10.1109/PMAPS53380.2022.9810580.

[18] Whei-Min Lin, Chin-Der Yang, Jia-Hong Lin and Ming-Tong Tsay, "A fault classification method by RBF neural network with OLS learning procedure," in IEEE Transactions on Power Delivery, vol. 16, no. 4, pp. 473-477, Oct. 2001, doi: 10.1109/61.956723.

[19] . N. H. Kothari, P. Tripathi, B. R. Bhalja and V. Pandya, "Support Vector Machine Based Fault Classification and Faulty Section identification Scheme in Thyristor Controlled Series Compensated Transmission Lines," 2020 IEEE-HYDCON, 2020, pp. 1-5, doi: 10.1109/HYDCON48903.2020.9242719.

[20] A. Jamehbozorg and S. M. Shahrtash, "A Decision Tree-Based Method for Fault Classification in Double-Circuit Transmission Lines," in IEEE Transactions on Power Delivery, vol. 25, no. 4, pp. 2184-2189, Oct. 2010, doi: 10.1109/TPWRD.2010.2050911.

[21] A. Harish, A. Prince and M. V. Jayan, "Fault Detection and Classification for Wide Area Backup Protection of Power Transmission Lines Using Weighted Extreme Learning Machine," in IEEE Access, vol. 10, pp. 82407-82417, 2022, doi:10.1109/ACCESS.2022.3196769.

[22] X. D. Wang, X. Gao, Y. M. Liu and Y. W. Wang, "WRC-SDT Based On-Line Detection Method for Offshore Wind Farm Transmission Line," in IEEE Access, vol. 8, pp. 53547-53560, 2020, doi: 10.1109/ACCESS.2020.2981294.

[23] Y. Ma and H. Duan, "Fault Line Selection Method for Small Cuttent Grounding Based on RBF-FCM," 2022 IEEE 10th Joint International Information Technology and Artificial Intelligence Conference (ITAIC), 2022, pp. 34-38, doi: 10.1109/ITAIC54216.2022.9836555.

[24] Robi Polikar."The Story of Wavelets", Iowa State University.

[25] J. W. Ballance and S. Goldberg, "Subsynchronous Resonance in Series Compensated Transmission Lines," in IEEE Transactions on Power Apparatus and Systems, vol. PAS-92, no. 5, pp. 1649-1658, Sept. 1973, doi: 10.1109/TPAS.1973.293713.

[26] K.R.Padiyar, Power System Dynamics, Stability Control, 2nd Edition,B.S. Publications, Hyderabad, 2002.